 \documentclass[preprint2]{aastex}

\newcommand{\VEV}[1]{\langle#1\rangle}

\shorttitle{The local curvature of the WMAP data.}
\shortauthors{Hansen,Cabella,Marinucci,Vittorio}

\begin{document}


\title{Asymmetries in the local curvature of the WMAP data.}

\author{Frode K. Hansen}
\affil{Dipartimento di Fisica, Universit\`a di Roma `Tor Vergata', Via della Ricerca Scientifica 1, I-00133 Roma, Italy}
\email{frodekh@roma2.infn.it}

\author{Paolo Cabella}
\affil{Dipartimento di Fisica, Universit\`a di Roma `Tor Vergata', Via della Ricerca Scientifica 1, I-00133 Roma, Italy}
\email{Paolo.Cabella@roma2.infn.it}

\author{Domenico Marinucci}
\affil{Dipartimento di Matematica, Universit\`a di Roma `Tor Vergata', Via della Ricerca Scientifica 1, I-00133 Roma, Italy}
\email{marinucc@mat.uniroma2.it}

\and

\author{Nicola Vittorio}
\affil{Dipartimento di Fisica, Universit\`a di Roma `Tor Vergata', Via della Ricerca Scientifica 1, I-00133 Roma, Italy}
\affil{INFN, Sezione di Roma 2, Via della Ricerca Scientifica 1, I-00133 Roma, Italy}

\begin{abstract}
We use the local curvature to investigate the possible existence of non-Gaussianity/asymmetry in the WMAP data. Considering the full sky we find results which are consistent with the Gaussian assumption. However, strong non-Gaussian features emerge when considering the northern and southern galactic hemisphere separately, particularly on scales between 1 and 5 degrees. Quite interestingly, the maximum non-Gaussianity is found for hemispheres centered near the ecliptic poles, which might suggest the presence of some systematic effect. The direction of the asymmetry seems consistent with the findings by Eriksen et al. 2004.
\end{abstract}

\keywords{ (cosmology:) cosmic microwave background --- cosmology: observations --- methods: data analysis ---  methods: statistical}   

\section{Introduction}
\label{sect:intro}

Many recent papers \citep{komatsu,naselsky,naselsky2,chiang,park,eriksen,naselsky3,coles,copi,vielva,magueijo,eriksen2} have tested the Gaussianity of the cosmic microwave background (CMB) on the WMAP data. The most striking result from some of these papers is the possible existence of an asymmetry in the distribution of the CMB fluctuations on different hemispheres. For instance \citep{eriksen2} find strong non-Gaussianity, as measured by the genus, on the northern galactic hemisphere; \citep{vielva} detect, by using a wavelet analysis, a distinct non-Gaussian feature in the southern hemisphere; similarly results by \citep{copi,eriksen,park} seem inconsistent with the assumption that CMB fluctuations are isotropic/Gaussian on various scales. The physical implications of these findings are still unclear. In this letter we aim at a further investigation on this issue by using a different method; namely the local curvature approach advocated by \citep{dore}, which we extend from the flat sky approximation to the full sky setting \citep{paolo}.\\
The plan of the paper is as follows; in section (\ref{sect:curv}) we briefly describe the local curvature formalism on the sphere, section (\ref{sect:method}) outlines our method and section (\ref{sect:results}) collects the results, which we discuss in section (\ref{sect:concl}).

\section{Local curvature on the sphere}
\label{sect:curv}

Similarly to \citep{dore} we classify the points of the renormalized map:
\begin{equation}
T(\theta,\phi)\rightarrow \frac{T(\theta,\phi)-\VEV{T}}{\sigma(T)} 
\end{equation}    
in three types:
\begin{itemize}
\item \emph{ hills} where the eigenvalues of the Hessian  are both positive.
\item  \emph{lakes}  where the eigenvalues of the Hessian  are both negative.
\item \emph{saddles} where the eigenvalues  are of opposite sign.
\end{itemize}

We can hence probe non-Gaussianity by evaluating the proportions of hills, lakes and saddles above a certain threshold $\nu$ in the normalized map on a given part of the sky (see \citep{dore}). The Hessian on the sphere is calculated using the covariant second derivatives \citep{schmalzing,eriksen2}. The derivatives are calculated in spherical harmonic space. More details can be found in \citep{paolo}.

\section{The method}
\label{sect:method}
For the analysis performed in this Letter we used the publicly available WMAP data\footnote{available at http://lambda.gsfc.nasa.gov}. We used the co-added W+V maps and as a consistency check also the Q1+Q2 co-added map. The maps where co-added in the way described by \citep{bennett} and the unphysical monopole and dipole were removed using the $a_{\ell m}$ coupling kernel.  In order to lower the noise and to probe different scales, we have smoothed the maps with Gaussian beams of FWHM in the range $0^\circ$ to $15^\circ$. The hills, lakes and saddles were calculated on 512 simulated maps (in the case where the results are less interesting we have limited the number of simulations to 128 to save CPU time: this is clearly indicated in the tables) as well as on the WMAP data using the following procedure:
\begin{enumerate}
\item The map was multiplied with the Kp2 mask and normalized. In order to avoid problems with smoothing close to the boundaries of the mask, the point source holes were refilled, mirroring the area just outside of the holes. The galaxy was refilled mirroring the part just above (for the northern hemisphere) or below (southern hemisphere).
\item A spherical harmonic transform was applied and the map was smoothed with an appropriate Gaussian beam in spherical harmonic space. An inverse transform created the smoothed map; then the mask was applied again and a new normalization was performed.
\item We calculated the spherical harmonic coefficients of this smoothed and masked map in order to perform the derivatives in spherical harmonic space using a procedure similar to \citep{schmalzing,eriksen2}.
\item On the derivative maps we applied an extended Kp2 sky cut (in order to avoid problems with the derivatives close to the boundaries) and for some tests we extended the cut even further using the Kp0 or a $30^\circ\times2$ galactic cut. On these maps we counted the hills, lakes and saddles for different levels $\nu$ in the regions of interest.
\end{enumerate}

Tests on simulations have shown that the bias introduced by the refilling procedure for smoothing around the mask is minimal when compared to maps where the smoothing was performed on the full sky. Note also that in the final step, we use an extended mask to avoid the unstable derivatives close to the boundary. The extended mask was found using the derivative of the mask itself (a similar procedure was adopted in \citep{schmalzing} to remove the bias introduced by smoothing close to the boundaries of a sky cut). We included only the pixels in which the derivatives were above a certain level. The correct level was adjusted in such a way that the number count of hills, lakes and saddles were not significantly biased with respect to simulations with derivatives taken on the whole map (the original Kp2 mask removes $15\%$ of the sky whereas the extended mask removes $25\%$).

We have hence implemented a $\chi^2$ statistic:
\[
\chi^2_{x}=\sum_{\nu=-3\sigma}^{\nu=2\sigma} (x_\nu-\VEV{x_\nu})^2/\sigma_x(\nu)^2
\]
where $x_\nu$ represents the proportions of hills, lakes or saddles above a given threshold $\nu$ and $\sigma_x(\nu)=\sqrt{\VEV{x_\nu^2-\VEV{x_\nu}^2}}$. We have excluded the highest thresholds as there were very few pixels present and the statistics becomes very poor at these levels (see \citet{dore}).

\section{The results}
\label{sect:results}

We applied the test on the full sky W+V data using the $Kp2$ mask and a calibration of 128 simulations for each smoothing scale. The results are reported in table (\ref{tab:normal}), where the numbers represent the percentages of simulated maps with a higher $\chi^2$. We see that considering the full sky, the maps are consistent with the Gaussian hypothesis. However, motivated by the recent results where non-Gaussianities have been found only on the northern galactic hemisphere \citep{eriksen2} and with a different method only on the southern galactic hemisphere \citep{vielva}, we decided to implement the same test on these two hemispheres separately. The results are reported in table (\ref{tab:normal}) for the co-added W+V channel (512 simulations were used in the calibration, except when the percentages are reported as integers, in those cases 128 simulations were used). The northern hemisphere appears non-Gaussian at the $2-3\sigma$ level at scales between 1 and 5 degrees. The saddles are also non-Gaussian but only at the $1\sigma$ level and at the smallest scales. The southern hemisphere shows results consistent with the Gaussian hypothesis, except for the scales between 2 and 4 degrees.  Note that for these scales both hemispheres taken separately are non-Gaussian, whereas considering the full sphere, the results appear Gaussian. The reason is that the proportions of hills, lakes and saddles are exceptionally high in one hemisphere and exceptionally low in the other. For the scales between 8 and 15 degrees at which the northern hemisphere is Gaussian, the southern hemisphere produces results with particularly small departures from the Gaussian expected values; this may be due to a different angular power spectrum on the two hemispheres, yielding a lower variance than expected in the south (see \citep{eriksen}). As a consistency check, we performed the same analysis on the co-added Q1+Q2 map and found very similar results, thus the non-Gaussian features do not seem to depend on frequency channel. To investigate the possibility of a contaminating foreground, we have performed the same test on the W+V channel, using the Kp0 mask and a $30^\circ\times2$ galactic cut. The results are summarized in table (\ref{tab:cut}). The Kp0 mask does not lower the effect, however with the $30^\circ\times2$ galactic cut, the significance is reduced to the $1\sigma$ level. This does not imply that the feature which caused a significant non-Gaussian detection using a larger part of the sky is absent outside of this huge cut. We will see at the end of this section that the positional dependence of the $\chi^2$ does not change even when using the $30^\circ\times2$ galactic mask. Finally we also point out that the hill, lake and saddle counts are power spectrum dependent. However, trying small variations of the model power spectrum in the simulations does not seem to change the results significantly.  

In order to investigate the exact location of the non-Gaussian features, we implemented the tests on hemispheres centered at 164 different positions. In table (\ref{tab:cut}) we show the results on the maximally asymmetric directions. In each simulated map we estimated the $\chi^2$ on each of the 164 differently orientated hemispheres and we recorded the highest $\chi^2$ value. We did the same for the WMAP W+V map with the Kp2 sky cut. Taking into account the rotations, we see that the significance of the asymmetry is reduced, but still at the $2\sigma$ level for the 3 degree smoothing scale.

\begin{figure}
\includegraphics[angle=90,scale=.30]{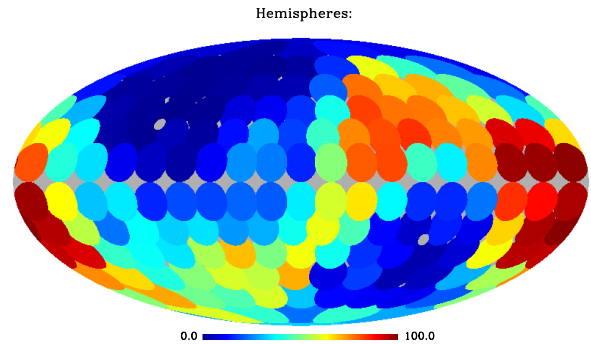}
\caption{The discs show the centeres of hemispheres on which the $\chi^2$ has been calculated for the lake counts. The numbers indicate the percentage of simulations with a higher $\chi^2$ for the given hemisphere. In the analysis, the extended Kp2 cut was used with a smoothing scale of $1^\circ$ FWHM. The two small dots indicate the position of the ecliptic poles. \label{fig:map}}
\end{figure}

In figure (\ref{fig:map}) we represent the dependence of non-Gaussian features from different directions for the 1 degree smoothing scale using the Kp2 cut. The discs represent the centers of the hemispheres; the numbers reported indicate the percentages of Gaussian simulations which yielded a higher $\chi^2$ value for the lakes for that given hemisphere. Note that there are two non-Gaussian areas, both close to the ecliptic poles, which could suggest the presence of a systematic effect at this scale. For the 3 degree smoothing scale (not shown), the map appears very similar, the main difference being that the two non-Gaussian 'poles' are now slightly closer to the galactic poles. However for the 5 degree smoothing scale, the non-Gaussian area in the southern hemisphere vanishes. Using the $30^\circ\times2$ galactic cut, the appearance of the map hardly changes, suggesting that the same features are still present, even if they are only at the $1\sigma$ significance level. Furthermore, the maps of the $\chi^2$ for the hills have a very similar appearance to those of the lakes; also, using the co-added Q1+Q2 channels, the same structure emerges. Figure (\ref{fig:map}) is strikingly similar to an analogous one in \citep{eriksen} which shows the distribution of the power spectrum and the three-point correlations function on the sky.

\section{Conclusions}
\label{sect:concl}

In this letter we have further investigated the possible existence of non-Gaussianity/asymmetry in the WMAP data. Considering the full sky data (using an extended Kp2 galaxy cut and point source mask) we find results which are consistent with the Gaussian assumption. However, strong non-Gaussian features (2-3$\sigma$) emerge when considering the northern and southern galactic hemisphere separately, particularly on scales between 1 and 5 degrees where the proportions of hills and lakes are extermely high on one hemisphere and extremely low on the other. Quite interestingly, the maximum asymmetry is found for hemispheres centered close to the ecliptic poles, which could suggest the presence of a systematic effect. For the 5 to 10 degree FWHM smoothing scale, the southern hemisphere yields results so close to the Gaussian expected values to suspect overestimation of the variance in this part of the sky. The position of the non-Gaussianity on the sky seems very consistent with the asymmetric distribution of the power spectrum and the three-point correlation functions found by \citep{eriksen}. 

\begin{acknowledgments}

We acknowledge the use of the Healpix package\footnote{http://www.eso.org/science/healpix/}. FKH acknowledges financial support from the CMBNET Research Training Network. This research used resources of the National Energy Research Scientific Computing Center, which is supported by the Office of Science of the U.S. Department of Energy under Contract No. DE-AC03-76SF00098. We also acknowledge the use of the supercomputers at Cineca (Bologna).

\end{acknowledgments}

\begin{deluxetable}{lccc}
\tabletypesize{\small}
\tablecaption{Results \label{tab:normal}}
\tablewidth{0pt}
\tablehead{
\colhead{FWHM} & \colhead{hills} & \colhead{lakes} & \colhead{saddles}} 
\startdata
\multicolumn{4}{l}{\it full sky (Kp2)}\\
\hline
no sm.& 72     & 56 & 63  \\
$0.5^\circ$& 2 & 45 & 1  \\
$1^\circ$& 39  & 39 & 40  \\
$3^\circ$ &24  & 16 & 21  \\
$5^\circ$ &20  & 29 & 62  \\
$8^\circ$ &48  & 41 & 41  \\
$10^\circ$ &56 & 83 & 55  \\
$15^\circ$& 34 & 76 & 26  \\
\multicolumn{4}{l}{\it northern gal. hemisphere (Kp2)}\\
\hline
no sm.&3        &  8   & 16   \\     
$1^\circ$& 0.8   & 6.3  & 10.7 \\
$2^\circ$& 1     & 4    & 5 \\
$3^\circ$&0.8    & 1.4  & 13.5 \\ 
$4^\circ$&2      & 1    & 53 \\ 
$5^\circ$&1.4    & 1.8  & 60.9 \\  
$8^\circ$& 10.4  & 18.9 & 49.4 \\  
$10^\circ$& 13   & 34   & 37   \\ 
\multicolumn{4}{l}{\it southern gal. hemisphere (Kp2)}\\
\hline
no sm.& 28       &  40  &98    \\
$1^\circ$& 28.1  & 37.5 &75.0  \\
$2^\circ$& 2     & 20   & 63 \\
$3^\circ$& 2.5   & 3.9  &79.1  \\
$4^\circ$&25     & 2    & 23 \\ 
$5^\circ$& 70.1  & 32.6 &56.3  \\
$8^\circ$& 92.0  & 34.6 &78.3  \\
$10^\circ$& 91   &   83 &99    \\
\enddata

\end{deluxetable}

\begin{deluxetable}{lccc}
\tabletypesize{\small}
\tablecaption{Results \label{tab:cut}}
\tablewidth{0pt}
\tablehead{
\colhead{FWHM} & \colhead{hills} & \colhead{lakes} & \colhead{saddles}} 
\startdata
\multicolumn{4}{l}{\it northern gal. hemisphere (Kp0)}\\
\hline
$1^\circ$& 1.0  &7.8 & 16.8 \\
$3^\circ$& 1.6  &1.2 &20.9  \\
$5^\circ$& 3.1  &1.2 &46.3  \\
$8^\circ$&  13  &15  &50    \\
\multicolumn{4}{l}{\it southern gal. hemisphere (Kp0)}\\
\hline
$1^\circ$& 32.0 &50.4  &73.2  \\
$3^\circ$& 2.9  & 7.8  &80.9  \\
$5^\circ$& 58.4 & 50.0 & 93.2 \\
$8^\circ$&  77  &72    &91    \\
\multicolumn{4}{l}{\it northern gal. hemisphere ($30^\circ\times2$ gal. cut) }\\
\hline
$1^\circ$& 13   &30  &38    \\
$3^\circ$& 14.6 &9.6 &41.0  \\
$5^\circ$& 14.5 &7.4 &54.5  \\
$8^\circ$& 23   & 45 & 45   \\
\multicolumn{4}{l}{\it southern gal. hemisphere ($30^\circ\times2$ gal. cut) }\\
\hline
$1^\circ$& 49   &65    &94    \\
$3^\circ$& 8.6  &22.1  &89.0  \\
$5^\circ$& 68.8 &74.4  &100.0 \\
$8^\circ$& 68   &  84  &91    \\
\multicolumn{4}{l}{\it maximally non-Gaussian hemisphere (Kp2)}\\
\hline
$1^\circ$&6.1 &13.2  & 31.3 \\
$3^\circ$&5.7 &4.5   & 26.2 \\
$5^\circ$&10.4&9.4   & 59.8 \\
\enddata

\end{deluxetable}

\

\end{document}